\documentclass[useAMS,usenatbib]{mn2e}
\usepackage{hyperref}
\usepackage{graphicx}
\usepackage{amssymb}
\usepackage[usenames,dvipsnames]{color}
\newcommand\herschel{\textit{Herschel}}
\newcommand\microns{$\mu$m}
\newcommand\kms{\mbox{km\,s$^{-1}$}}
\newcommand{\hatlas}[1]{SDP.#1}
\newcommand{\lines}[1]{$L_{\mathrm{#1}}$}

\newcommand{\AIPS}{{$\cal AIPS\/$}}
\newcommand{\oiii}{[O\,{\sc iii}]}
\newcommand{\oi}{[O\,{\sc i}]}
\newcommand{\cii}{[C\,{\sc ii}]}
\newcommand{\nii}{[N\,{\sc ii}]}
\newcommand{\hii}{\mbox{H\,{\sc ii}}}

\title[ISM of H-ATLAS submm galaxies]{Physical conditions of the interstellar medium of high-redshift, strongly lensed submillimetre galaxies from the \herschel-ATLAS\thanks{{\it Herschel} is an ESA space observatory with science instruments provided by European-led Principal Investigator consortia and with important participation from NASA.}}
\author[Valtchanov et al.]
{
I. Valtchanov$^{1}$\thanks{E-mail:ivaltchanov@sciops.esa.int},
J. Virdee$^{2,3}$,
R.\,J.\ Ivison$^{4,5}$, 
B. Swinyard$^{2,6}$,
P. van der Werf$^{7,5}$,\newauthor
D. Rigopoulou$^{2,3}$,
E. da Cunha$^{8,9}$,
R. Lupu$^{10}$,
D. J. Benford$^{11}$, 
D. Riechers$^{12}$,\newauthor 
Ian Smail$^{13}$,
M. Jarvis$^{14}$,
C. Pearson$^{2}$,
H. Gomez$^{15}$,
R. Hopwood$^{16}$,
B. Altieri$^{1}$,\newauthor
M. Birkinshaw$^{17}$,
D. Coia$^{1}$,
L. Conversi$^{1}$,
A. Cooray$^{18}$,
G. De Zotti$^{19,20}$,
L. Dunne$^{21}$,\newauthor
D. Frayer$^{22}$,
L. Leeuw$^{23,24}$,
A. Marston$^{1}$, 
M. Negrello$^{25}$,
M. Sanchez Portal$^{1}$,\newauthor
D. Scott$^{26}$,
M.\,A. Thompson$^{21}$,
M. Vaccari$^{27}$,
M. Baes$^{28}$,
D. Clements$^{16}$,\newauthor
M.\,J. Micha{\l}owski$^{5}$,
H. Dannerbauer$^{29}$,
S. Serjeant$^{25}$,
R. Auld$^{15}$,
S. Buttiglione$^{27}$,\newauthor
A. Cava$^{30}$,
A. Dariush$^{15,31}$,
S. Dye$^{15}$,
S. Eales$^{15}$,
J. Fritz$^{28}$,
E. Ibar$^{4}$,
S. Maddox$^{21}$,\newauthor
E. Pascale$^{15}$,
M. Pohlen$^{15}$,
E. Rigby$^{21}$,
G. Rodighiero$^{27}$, 
D.\,J.\,B. Smith$^{21}$,
P. Temi$^{32}$,\newauthor
J. Carpenter$^{12}$,
A. Bolatto$^{33}$,
M. Gurwell$^{34}$,
J.D. Vieira$^{12}$\\
$^{1}$European Space Astronomy Centre, Herschel Science Centre, ESA, 28691 Villanueva de la Ca\~nada, Spain\\
$^{2}$RAL Space, STFC-Rutherford Appleton Laboratory, Harwell Campus, Chilton, Didcot, Oxon, OX11 0QX, UK \\
$^{3}$Oxford Astrophysics, Denys Wilkinson Building, University of Oxford, Keble Road, Oxford, OX1 3RH, UK\\
$^{4}$UK Astronomy Technology Centre, Royal Observatory, Blackford Hill, Edinburgh EH9 3HJ, UK\\
$^{5}$SUPA, Institute for Astronomy, University of Edinburgh, Blackford Hill, Edinburgh EH9 3HJ, UK\\
$^{6}$Department of Physics and Astronomy, University College London, Gower Street, London, WC1E 6BT, UK\\
$^{7}$Leiden Observatory, Leiden University, P.O. Box 9513, 2300 RA Leiden, The Netherlands\\
$^{8}$Department of Physics, University of Crete, 71003 Heraklion, Greece\\
$^{9}$Max-Planck-Institut fur Astronomie, Konigstuhl 17, 69117 Heidelberg, Germany\\
$^{10}$Department of Physics and Astronomy, University of Pennsylvania, Philadelphia, PA19104, USA\\
$^{11}$Observational Cosmology Laboratory (Code 665), NASA Goddard Space Flight Center, Greenbelt, MD 20771, USA\\
$^{12}$California Institute of Technology, 1200 E. California Blvd., MC 249-17, Pasadena, CA 91125, USA\\
$^{13}$Institute for Computational Cosmology, Department of Physics, Durham University, Durham DH1 3LE, UK\\
$^{14}$Centre for Astrophysics Research, Science and Technology Research Centre, University of Hertfordshire, Herts AL10 9AB, UK\\
$^{15}$School of Physics and Astronomy, Cardiff University, The Parade, Cardiff, CF24 3AA, UK \\
$^{16}$Astrophysics Group, Imperial College, Blackett Laboratory, Prince Consort Road, London SW7 2AZ, UK\\
$^{17}$HH Wills Physics Laboratory, University of Bristol, Tyndall Avenue, Bristol BS8 1TL, UK\\
$^{18}$Department of Physics \& Astronomy, University of California, Irvine, CA 92697, USA\\
$^{19}$INAF, Osservatorio Astronomico di Padova, Vicolo Osservatorio 5, I-35122 Padova, Italy\\
$^{20}$Scuola Internazionale Superiore de Studi Avanzati, Via Bonomea 265, I-34136 Trieste, Italy\\
$^{21}$School of Physics and Astronomy, University of Nottingham, University Park, Nottingham NG7 2RD, UK\\
$^{22}$National Radio Astronomy Observatory, PO Box 2, Green Bank, WV 24944, USA  \\
$^{23}$Physics Department, University of Johannesburg, PO Box 524, Auckland Park 2006, South Africa\\
$^{24}$SETI Institute, 515 N. Whishman Avenue, Mountain View, CA 94043, USA\\
$^{25}$Department of Physics \& Astronomy, The Open University, Milton Keynes, MK7 6AA, UK\\
$^{26}$Department of Physics \& Astronomy, University of British Columbia, 6224 Agricultural Road, Vancouver, BC V6T 1Z1, Canada\\
$^{27}$Dipartimento di Astronomia, Universit\`a di Padova, vicolo Osservatorio 3, I-35122 Padova, Italy\\
$^{28}$Sterrenkundig Observatorium, Universiteit Gent, Krijgslaan 281 S9, B-9000 Gent, Belgium\\
$^{29}$DAPNIA/Service d'Astrophysique, CEA Saclay, Orme des Merisiers, 91191 Gif-sur-Yvette, France\\
$^{30}$Departamento de Astrof\'{\i}sica, Facultad de CC. F\'{\i}sicas, Universidad Complutense de Madrid, E-28040 Madrid, Spain\\
$^{31}$School of Astronomy, Institute for Research in Fundamental Sciences (IPM), PO Box 19395-5746, Tehran, Iran\\
$^{32}$Astrophysics Branch, NASA Ames Research Center, Mail Stop 245-6, Moffet Field, CA 94035, USA\\
$^{33}$Department of Astronomy, University of Maryland, College Park, MD 20742, USA\\
$^{34}$Harvard-Smithsonian Center for Astrophysics, Cambridge, MA 02138, USA
}

\begin{document}

\date{Accepted \today. Received \today; in original form \today}

\pagerange{\pageref{firstpage}--\pageref{lastpage}} \pubyear{2010}

\maketitle

\label{firstpage}

\begin{abstract}
We present \herschel-SPIRE Fourier Transform Spectrometer (FTS) and radio follow-up observations of two \herschel-ATLAS (H-ATLAS) detected strongly lensed distant galaxies. In one of the targeted galaxies H-ATLAS J090311.6$+$003906 (\hatlas{81}) we detect \oiii\,88\,\microns\ and  \cii\,158\,\microns\ lines at a signal-to-noise ratio of $\sim5$.  We do not have any positive line identification in the other fainter target H-ATLAS J091305.0$-$005343 (\hatlas{130}). Currently \hatlas{81} is the faintest sub-mm galaxy with positive line detections with the FTS, with continuum flux just below 200 mJy in the 200-600 \microns\ wavelength range. The derived redshift of \hatlas{81} from the two detections is $z=3.043\pm 0.012$, in agreement with ground-based CO measurements. This is the first detection by \herschel\ of the \oiii\,88 \microns\ line in a galaxy at redshift higher than 0.05. Comparing the observed lines and line ratios with a grid of photo-dissociation region (PDR) models with different physical conditions, we derive the PDR cloud density $n \approx 2000$ cm$^{-3}$ and the far-UV ionizing radiation field $G_0 \approx 200$ (in units of the Habing field -- the local Galactic interstellar radiation field of $1.6\times10^{-6}$ W m$^{-2}$). Using the CO derived molecular mass and the PDR properties we estimate the effective radius of the emitting region to be 500-700 pc. These characteristics are typical for star-forming, high redshift galaxies. The radio observations indicate that \hatlas{81} deviates significantly from the local FIR/radio correlation, which hints that some fraction of the radio emission is coming from an AGN. The constraints on the source size from millimiter-wave observations put a very conservative upper limit of the possible AGN contribution to less than 33\%. These indications, together with the high \oiii/FIR ratio and the upper limit of \oi\,63\,\microns/\cii\,158\,\microns\ suggest that some fraction of the ionizing radiation is likely to originate from an AGN. 
\end{abstract}

\begin{keywords}
galaxies: evolution --- infrared: galaxies --- submillimetre: ISM --- radio continuum: galaxies --- galaxies: individual (\hatlas{81}: H-ATLAS J090311.6$+$003906, \hatlas{130}: H-ATLAS J091305.0$-$005343)
\end{keywords}

\section{Introduction}
The detection of large numbers of high-redshift, optically obscured, massive star-forming galaxies by ground-based sub-mm instruments such as Submmilimetre Common-User Bolometer Array (SCUBA) and Max Planck Millimetre Bolometer (MAMBO) was a surprise and a major discovery (e.g. \citealt{smail97}). These submm galaxies (SMGs) present amongst the most difficult obstacles for models of galaxy formation and evolution (e.g. \citealt{baugh05}). The SMGs are a subset of the dusty star-forming galaxies that account for more than half of the far-infrared background \citep{devlin09}. Their star-formation rates are orders of magnitudes higher than most galaxies in the local Universe and they are considered the precursors of today's most massive elliptical or bulge galaxies \citep{lilly99,swinbank06}. 

Due to the fact that more than half of the light from the SMGs is absorbed and reprocessed by dust, it is exceedingly important to study these sources at their peak emission, which occurs in the far infrared (FIR). Prior to the \herschel\  \citep{pilbratt10}, ground based and balloon-borne studies of SMGs were limited to small fields and produced modest samples detected at low significance. This situation is changing dramatically with the number of wide-field surveys with \herschel\ Space Observatory, such as the \herschel\ Astrophysical Terahertz Large Area Survey (H-ATLAS). H-ATLAS is the largest \herschel\ Open Time Key Programme \citep{eales10}, and will map 550 deg$^2$ in six extragalactic fields in five photometric bands, using in parallel mode both the PACS \citep{pacs} and the SPIRE \citep{spire} instruments. One of the most interesting results of H-ATLAS will be the detection of a significant number of gravitationally lensed objects in the distant ($z>2$) Universe. This is possible in such a shallow but large area survey because of the combined effects of a strong negative $K$-correction and steep number counts in the SPIRE 500-\microns\ waveband (e.g. \citealt{blain96}). Objects brighter than 100\,mJy at 500\,\microns\ will be a mixture of lensed high-redshift galaxies, nearby galaxies and flat-spectrum radio sources \citep{dezotti05,negrello07,eales10} and current models predict that H-ATLAS will detect some $\sim 350$ strongly lensed objects at 500\,\microns\ at redshifts $>1$ \citep{negrello07}.

High-redshift, bright SMGs open a new possibility for the detailed study of the physical conditions of the interstellar medium, at rest-frame far-infrared. Atomic emission lines in this range provide the major coolants for neutral gas exposed to the far-ultraviolet (FUV) ionizing flux of nearby early-type stars. Star-forming molecular clouds where the physical processes are dominated by FUV photons  are called photo-dissociation regions (PDRs, e.g. \citealt{hollenbach97}). FIR lines are not strongly affected by extinction in these regions, and so provide an excellent probe of the embedded stellar radiation fields and their effects on the physical conditions of the neutral gas. The majority of their cooling occurs via atomic fine structure lines of \cii, \oi, \oiii\  \citep{brauher08} and the molecular rotational $^{12}\mathrm{CO}$ lines \citep{kramer04}.  There have been several pre-\herschel\ efforts to study fine structure lines for nearby as well as higher redshift galaxies, such as \citet{stacey91,stacey10,malhotra01,maiolino05,maiolino09,brauher08,walter09a,walter09b,wagg10,ferkinhoff10,vasta10,hailey10}, as well as molecular (CO) studies of SMGs, e.g. \citet{greve05,tacconi06,tacconi08,daddi09a,daddi09b,riechers10,danielson11}. \herschel, however opens up a new frontier with its highly sensitive spectrometers, especially for observations of lines like \cii\  at 158 \microns\ at redshifts up to $\sim 3$.

During \herschel's Science Demonstration Phase, one of the H-ATLAS fields, GAMA9, was observed, covering $\sim 4\times 4\ \deg^2$ (see \citealt{smith11} and \citealt{rigby11} for the H-ATLAS data in this field). The processing of the \herschel\ observations is presented in \citet{ibar10b} for the PACS and in \citet{pascale11} for the SPIRE observations. We have identified 11 sources with flux density at 500 \microns\ $S_{500} \geq 100$\,mJy  and the results of the analysis of these candidate lensed sources, from follow-up observations using facilities both on the ground and in space, are presented in \citet{negrello10, lupu10, frayer10, hopwood10}.

\begin{table*}
\caption{Positions and photometry measurements of \hatlas{81} and \hatlas{130}. The quoted positions are those determined using IRAM PdBI at $0.62\arcsec \times 0.30\arcsec$ beam resolution (Neri et al. in preparation), thus accurate to $<1\arcsec$. The redshifts and the SPIRE photometry are taken from \citet{negrello10}, while the PACS 160 \microns\ measurements are from this work. The upper limits at 70 \microns\ are $1\sigma$.}
\begin{center}
\begin{tabular}{llllllllllll}
\hline\hline
Target ID & \multicolumn{1}{c}{Redshift} & \multicolumn{1}{c}{RA} & \multicolumn{1}{c}{Dec.} & \multicolumn{1}{c}{$S_{70}$} & \multicolumn{1}{c}{$S_{160}$} & \multicolumn{1}{c}{$S_{250}$} & \multicolumn{1}{c}{$S_{350}$} & \multicolumn{1}{c}{$S_{500}$} \\
 &  & \multicolumn{1}{c}{(J2000)} & \multicolumn{1}{c}{(J2000)} & \multicolumn{1}{c}{(mJy)} & \multicolumn{1}{c}{(mJy)} & \multicolumn{1}{c}{(mJy)} & \multicolumn{1}{c}{(mJy)} & \multicolumn{1}{c}{(mJy)}\\
\hline
\hatlas{81} & $3.042 \pm 0.001$ & 09:03:11.6 & $+$00:39:06.7 & $< 8$ & $51.4 \pm 5.0$ & $129.0 \pm 6.6$ & $181.9 \pm 7.3$ & $165.9 \pm 9.3$\\
\hatlas{130} & $2.625 \pm 0.001$ & 09:13:05.3 & $-$00:53:42.8 & $< 9$ & $45.3 \pm 7.7$ & $105.0 \pm 6.5$ & $127.6 \pm 7.2$ & $107.5 \pm 9.0$\\
\hline\hline
\end{tabular}
\label{tab1}
\end{center}
\end{table*}

The subject of this paper is to present the SPIRE Fourier-Transform  Spectrometer (FTS) and radio observations and report on the first results from the far-IR spectroscopy for two targets from \citet{negrello10}: H-ATLAS J090311.6$+$003906 designated \hatlas{81} and H-ATLAS J091305.0$-$005343 designated \hatlas{130}. Their positions, measured redshift and photometric properties in the FIR are shown in Table~\ref{tab1}. Each of these two targets was initially associated with an elliptical galaxy, yet based on their Spectral Energy Distribution (SED), peaking at $\sim 350$ \microns, it was clear that the foreground galaxy acts as a lens for a background source. Subsequent follow-up using Caltech Submmilimeter Observatory/Z-Spec and Green Bank Telescope/Zpectrometer instruments detected several CO lines at high signal-to-noise ratio (SNR) and secured a spectroscopic redshift of $3.042 \pm 0.001$ for \hatlas{81} and $2.625 \pm 0.001$ for \hatlas{130} (see \citealt{negrello10,lupu10,frayer10} for more details). 

The feasibility of the proposed FTS follow-up was well demonstrated during SPIRE's Performance and Verification phase. One bright, lensed, high-redshift SMG, SMM\,J2135$-$0102 at $z=2.3$ \citep{swinbank10}, was observed with the SPIRE FTS, achieving a detection of the \cii\  158\microns\ line at $4.3\sigma$ (line flux of \mbox{$(1.7\pm0.4) \times 10^{-17}$ W m$^{-2}$}) as well as tentative detections of \oi\ at 145.5\microns\ and \nii\ at 122.1\microns\ \citep{ivison10d}.

The structure of the paper is as follows: in Section~\ref{sec_obs} we present the FTS and radio observations and the FTS observing strategy. Some of the characteristics of the environment (telescope and instrument) during the FTS observations, relevant for the subsequent data processing are also presented. The FTS data processing, with some details on the deviations from the standard \herschel-SPIRE FTS data processing, is described in Section~\ref{sec_dp}. The results are presented in Section~\ref{sec_res} and summarised in the conclusions (Section~\ref{sec_conclusions}).

For all cosmological dependent parameters we use $h = 0.73$, $\Omega_m = 0.24$ and flat geometry, i.e. $\Omega_{\Lambda} = 1 - \Omega_m = 0.76$ \citep{cosmo}. With these parameters the luminosity distances to redshifts 2.625 and 3.042 are 22.06 and 26.41 Gpc, respectively.

\section{Observations}
\label{sec_obs}

\subsection{SPIRE FTS}
\label{sec_fts_obs}

The FTS observations were carried out on June 1st 2010 (\herschel\ observing day number 383). We targeted the two H-ATLAS lensed candidates, \hatlas{81} and \hatlas{130} and a nearby (offset at $\sim 2\arcmin$) dark sky region, devoid of sources and chosen using the available SPIRE maps. The three observations were done in a sequence: \hatlas{81} then the dark sky and then \hatlas{130}, each one with exactly the same SPIRE-FTS observing template: 100 repetitions using single pointing mode, sparse spatial sampling and high spectral resolution. The total time for each observation, including overheads, was 3$^\mathrm{h}$52.2$^\mathrm{m}$, with 66.6 s on-source time per FTS scan (one repetition consists of one forward and one reverse scan, i.e. two scans per repetition). 

The measured absolute pointing error of \herschel\ is $2\arcsec$ (68\%, \citealt{pilbratt10}) -- this is much smaller than the smallest FTS beam of $\sim 17\arcsec$ at 200 \microns\ \citep{swinyard10}. The pointing stability of \herschel\ during the observation was excellent: during the 4 hours of spectral scans the telescope jitter stayed within $0.14\arcsec$ (68\%) from the target position. The largest deviation was $0.54\arcsec$.

The main source of emission in the FTS bands is the \herschel\ telescope itself, although this is a constant background as the telescope temperature does not change significantly on the time scale of our observations. The emission of the telescope ranges from 100 Jy to 1000 Jy in the FTS bands (Swinyard et al., in prepration). Targets like \hatlas{81} and \hatlas{130}, with continuum levels around or below 200 mJy are just a tiny fraction of this signal. That is why it is very important to have an off-source reference dark sky measurement, performed as close in time as possible to the target observation, in order to be used to subtract the telescope and instrument contribution from the source spectrum (see Section~\ref{sec_dp}).

By design, the FTS also has a second input port that sees a spectrometer calibration source (SCAL) and this is the second source of significant thermal emission. Regardless of the fact that SCAL is not used and the lamp is turned off (see \citealt{OM}\footnote{The SPIRE Observers' Manual is available from the Herschel Science Centre:\\ \url{http://herschel.esac.esa.int/Docs/SPIRE/pdf/spire\_om.pdf}}), it stays at the instrument enclosure temperature and emits thermally in the two FTS bands. This instrument contribution is also removed with the help of the dark sky observation. As \hatlas{130} is about 30\% fainter than \hatlas{81} at 350 \microns\ and at a different redshift (see Table~\ref{tab1}), we have added this observation as a second dark sky measurement in order to decrease the noise of the reference subtraction for \hatlas{81}.

During the observations the primary mirror temperature was on average $88.20 \pm 0.06$ K, the secondary mirror temperature stayed at $84.290 \pm 0.004$ K with variations at the mK level. The instrument enclosure temperature during the \hatlas{81} observation was 4.70-4.75 K, during the off-source dark sky measurement it was 15 mK colder and 22 mK colder during the \hatlas{130} observation. Such seemingly small differences in the instrument and telescope temperatures, even at a level of a few mK, during the dark sky and the target measurements, can lead to overall additive effect on the continuum level. Following \citet{ivison10d}, we have used the three SPIRE photometry points at 250, 350 and 500 \microns\ in order to recover the continuum level (see Section~\ref{sec_dp}).

%\begin{figure}
%\includegraphics[width=8cm]{images/SCAL-temp.jpg}
%\caption{SPIRE SCAL temperature during the observations. The top curve in blue, labeled 0x50004edb is for \hatlas{81}, the middle curve in cyan is for the dark sky measurement. THIS FIGURE WILL PROBABLY BE REMOVED IN THE FINAL VERSION.}
%\label{fig_scal}
%\end{figure}

\subsection{PACS photometry}
In addition to the SPIRE FTS spectral scans we did short PACS cross-linked mini-scan maps for each source at 70 and 160 \microns\ (PACS blue and red bands). This was necessary because \hatlas{81} was not bright enough to be detected in the PACS parallel mode scans, while \hatlas{130} was not in the region covered by PACS in the PACS/SPIRE parallel mode. The PACS mini-scan maps were processed with version 5.0 of the \herschel\ Interactive Processing Environment (HIPE, \citealt{hipe}) using a standard data reduction script applying a temporal high-pass filter to remove the 1/f noise after masking the source for the filtering. The data cube was projected with the \textit{photProject} task and the 2 scan directions were combined with the \textit{Mosaic} task. Unfortunately the chosen integration time of $\sim 15$ minutes was not sufficient to detect the two sources at 70 \microns, where the uncertainties in the SED were the highest. The PACS flux density of {both targets} was measured in an aperture of $10\arcsec$ radius, corrected for the encircled energy fraction (see the PACS Observers' Manual\footnote{The PACS Observers' Manual is available from the Herschel Science Centre:\\ \url{http://herschel.esac.esa.int/Docs/PACS/pdf/pacs\_om.pdf}}) and are reported in Table~\ref{tab1}. At 70 \microns\ we only report the $1\sigma$ upper limits.

%\begin{figure}
%\includegraphics[width=8cm]{images/id81-SED-photo-FTS.pdf}
%\caption{Spectral energy distribution of \hatlas{81}. The FTS high resolution spectrum is also shown. The PACS upper limit photometry points at 70 and 100 \microns\ are shown as open upside-down triangles. We also include observations at 880 \microns\ (SMA) and 1.2 mm (MAMBO/IRAM). Templates from \citet{rieke09}, \citet{ce01} and \citet{negrello10} are also shown for comparison. Note that we use \citet{negrello10} template to estimate $L_\mathrm{IR}$.}
%\label{fig_sed}
%\end{figure}
%

%
%
\subsection{Radio observations}
\label{evla}

\hatlas{81} was observed by the Extended Very Large Array (EVLA) for two 3-hr tracks during 2010 July 17--18 (programme AI143/10A-253), when the EVLA was in its most compact configuration (D), recording data every 1\,s and using 16 Ka-band antennas. Antenna pointing was checked at 5\,GHz. Much of the first track suffered from a correlator problem and was discarded. For \hatlas{81}, the $^{12}$CO $J\!=\!1\!-\!0$ line (CO(1-0) hereafter) is redshifted to 28.53\,GHz for $z=3.042$, a frequency at which just one of the two sub-band pairs of the EVLA's new Ka-band receivers could be utilised, yielding 128\,MHz of instantaneous, dual-polarisation bandwidth during the earliest shared-risk phase of EVLA commissioning with the new Wide band Interferometric Digital Architecture (WIDAR) correlator. For our observations at 8.40\,GHz, we obtained 256\,MHz of contiguous, dual-polarisation bandwidth.  In total, we recorded 2\,hr of data at 8.40\,GHz and a similar amount at 28.53\,GHz, resulting in an overall on-source integration time of $\sim 85$ minutes. 3C\,286 was used for absolute flux calibration, and the local calibrator, J0909+0121, was observed every 5\,min.

%\begin{table}
%\caption{Radio observations of \hatlas{81}. The EVLA observations at 8.5 GHz and PdBI and CARMA at 240 GHz are from this work. The 1.4 GHz photometry point is from the FIRST survey \citep{becker95}.}
%\begin{tabular}{lll}
%\hline\hline
%Target ID &  $S_\mathrm{8.4GHz}$ & $S_\mathrm{1.4GHz}$\\
% &  \multicolumn{1}{c}{($\mu$Jy)} & \multicolumn{1}{c}{($\mu$Jy)} \\
%\hline
%\hatlas{81} & $700 \pm 50$ & $950 \pm 150$ \\
%\hline\hline
%\end{tabular}
%\label{tab_radio}
%\end{table}

Editing and calibration were accomplished within \AIPS\ ({\sc 31dec10}). For 3C\,286, we used an appropriately scaled 22.5-GHz model to determine gain solutions; J0909+0121 was used to determine the spectral variation of the gain solutions (the ``bandpass''), after first removing atmospheric phase drifts on a timescale of 12\,s via self-calibration with a simple point-source model. We were able to employ standard \AIPS\ recipes throughout the data-reduction process, following the techniques employed by \citet{ivison10b}. The synthesized beam for the CO observations is $3.3\arcsec \times 2.4\arcsec$ and $\approx12\arcsec \times 8\arcsec$ for the continuum.

\begin{figure}
\includegraphics[width=8cm]{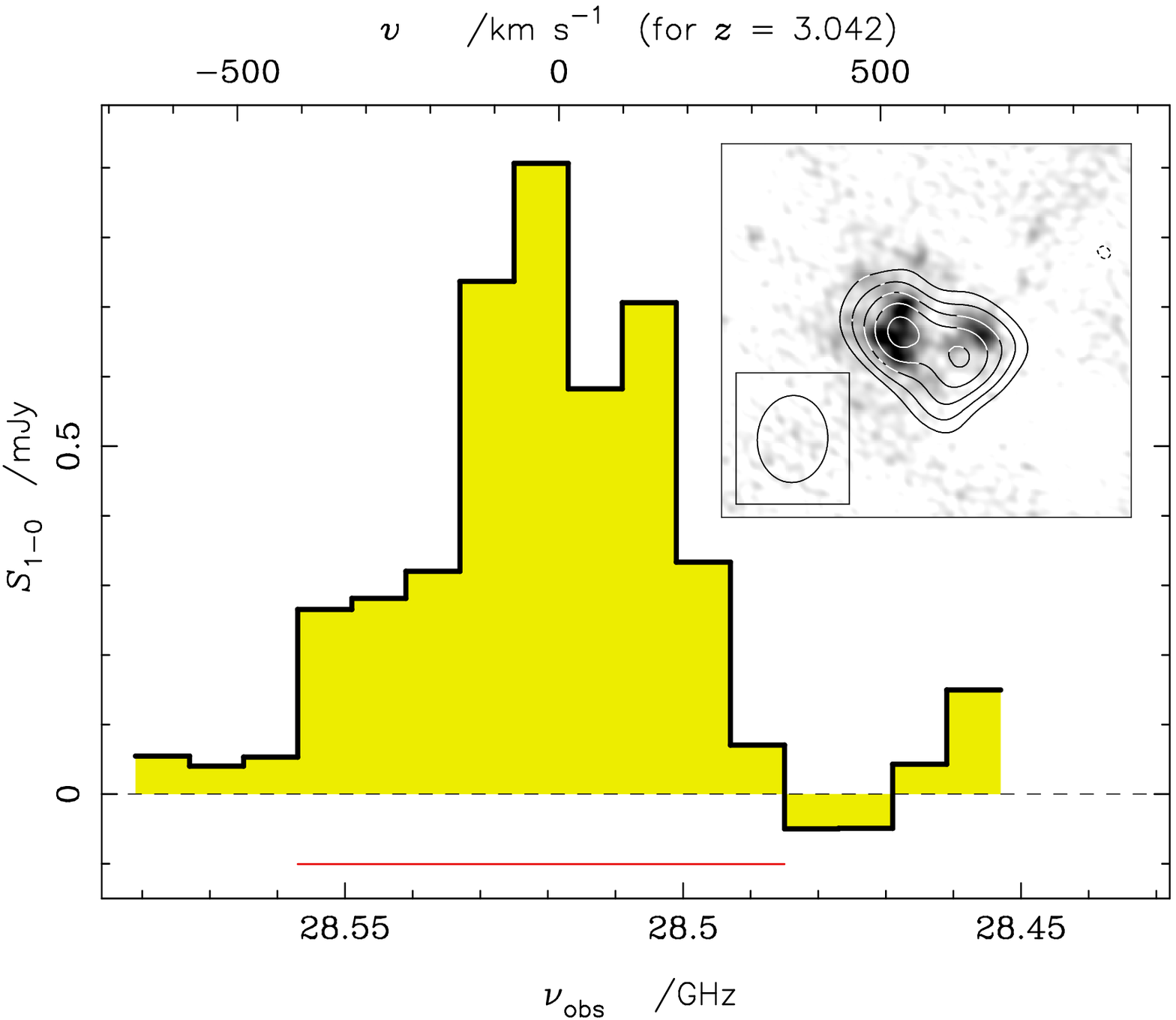}
\caption{Spectrum of $^{12}$CO $J=1-0$ line emission from \hatlas{81} from the Expanded Very Large Array. \emph{Inset:} Contours of the CO(1-0) emission, integrated across the velocity range $-$400 to +350\,\kms\ (indicated by the line below the spectrum, and where $v=0$\,km\,s$^{-1}$ at 28.51836\,GHz for $z=3.042$), which corresponds roughly to the FWZI of the line, overlaid upon a greyscale SMA 880 \microns\ continuum image (scaled from 0-8 mJy) presented by \citet{negrello10}. The synthesized beam for the CO observations is shown in the inset, bottom left ($3.3\arcsec \times 2.4\arcsec$), and the contours are plotted at $-$2.5, 2.5,3.5... 6.5$\times\sigma$, where the \textit{rms} noise level, $\sigma=100$\,$\mu$Jy\,beam$^{-1}$. The synthesized beam for the continuum observations is $\approx 12\arcsec \times 8\arcsec$ and the size of the image is $20\arcsec \times 20\arcsec$.}
\label{fig_co10_radio}
\end{figure}

Figure~\ref{fig_co10_radio} summarizes the result of the EVLA observations for \hatlas{81}. For CO(1-0), we measure an integrated line flux of $1.26\pm 0.20$\,Jy\,km\,s$^{-1}$, consistent with the GBT value of $1.11\pm 0.25$\,Jy\,km\,s$^{-1}$ \citep{frayer10}, the Full Width at Zero Intensity (FWZI) is $\sim750$ \kms. At 8.40\,GHz, we measure a total flux density of $S_{\rm 8.4GHz}=700\pm 50$\,$\mu$Jy. A contour plot of the integrated CO(1-0) emission is shown in the inset of Figure~\ref{fig_co10_radio}, overlaid on a greyscale 880 \microns\ continuum emission from the Submillimeter Array (SMA). 

\hatlas{81} was observed and detected with the Institut de Radioastronomie Millim\'etrique (IRAM) Plateau de Bure Interferometer (PdBI) at 240\,GHz and at resolution of $0.62\arcsec \times 0.30\arcsec$ on 2010 February 15th (Neri et al. in preparation). \hatlas{81} was also observed at higher resolution with the Combined Array for Research in Millimeter-wave Astronomy (CARMA) on 2010 February 22nd and 23rd (program cx306), when CARMA was in its most extended A configuration at 240 GHz. Due to previously arranged scheduling plans, the phase monitoring array was not deployed during these observations.  However, the phase \textit{rms} was less than 100\,\microns\ and the atmospheric opacity at 230 GHz (or $\tau_{\bf 230GHz}$) was less than 0.2 at zenith, allowing the observations to go ahead without the monitoring array.  The FWHM size of the test calibrator  was $0.19\arcsec \times 0.14\arcsec$, while the theoretical beam size is $0.17\arcsec \times 0.14\arcsec$.  This is pretty good agreement and suggests atmospheric ``seeing'' was not a big problem during the observations.

The CARMA observations of \hatlas{81} were in two $\sim 5$-hr tracks, the noise obtained from the first track being much higher than the second, due to the higher atmospheric opacity. Therefore, the two tracks were not coadded and only the results from second track that was observed in $\tau_{\bf 230GHz} = 0.1$ are presented here. The data were reduced with and without a gaussian taper of $0.3\arcsec$. The taper reduces the resolution (and sensitivity), but may pick out low-level emission. \hatlas{81} was not detected in either case, and the 240\,GHz respective resolutions and sensitivities reached with CARMA were $0.17\arcsec \times 0.14\arcsec$ and 0.67\,mJy/beam in the no-taper and $0.30\arcsec \times 0.26\arcsec$ and 1.1\,mJy/beam in the $0.3\arcsec$-taper maps.

\hatlas{81} is also present in the FIRST survey \citep{becker95} at 1.4 GHz with $S_\mathrm{1.4GHz} = 950 \pm 150\ \mu$m.

\section{SPIRE FTS data processing}
\label{sec_dp}

%The FTS processing is extremely challenging for such faint targets. Thanks to our optimal observing strategy (see the previous section) we were able to get the best possible outcome for the resulting spectra.

In this section we briefly describe our data processing approach and underline where it deviates from the standard SPIRE FTS data processing. Detailed description of the standard FTS pipeline can be consulted in the SPIRE Data Reduction Guide.\footnote{Available at \url{http://herschel.esac.esa.int}}

Each of the three observations, two targets and one reference dark/off-target, are processed separately out to the final spectrum. The forward and the reverse scans are kept separated throughout the whole processing. We consider only the central detectors of the two SPIRE FTS short-wavelength (SSW) and long-wavelength (SLW) bolometer arrays. Note that the off-axis bolometers are of a limited use as the optical path and the standing-wave (fringe) pattern in the interferograms are very different. The fringes are at high optical path differences and consequently they have the undesirable effect {of adding a fixed pattern noise that affects the high resolution part of the spectrum.}

\begin{figure*}
\includegraphics[width=8cm]{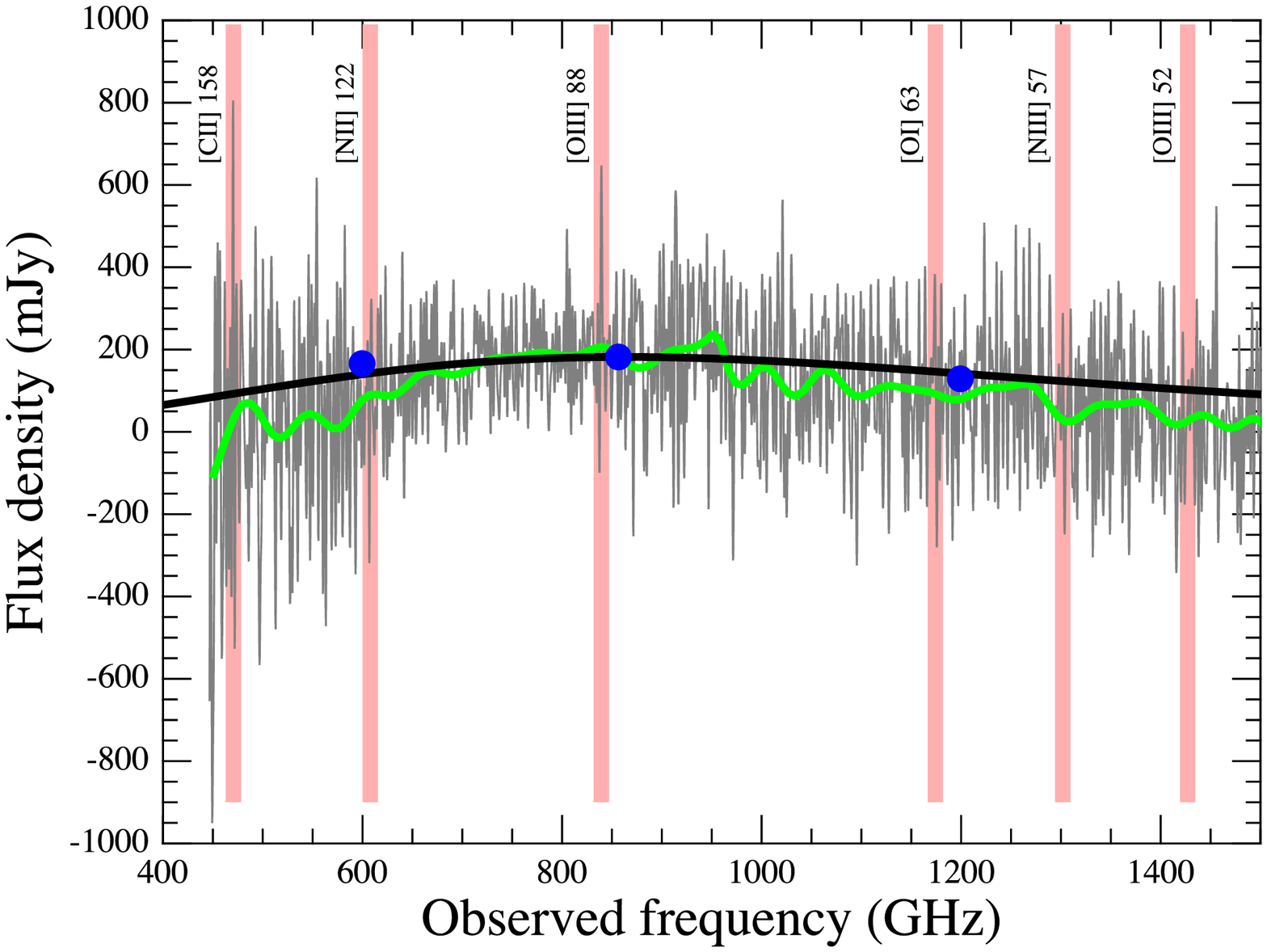}
\includegraphics[width=8cm]{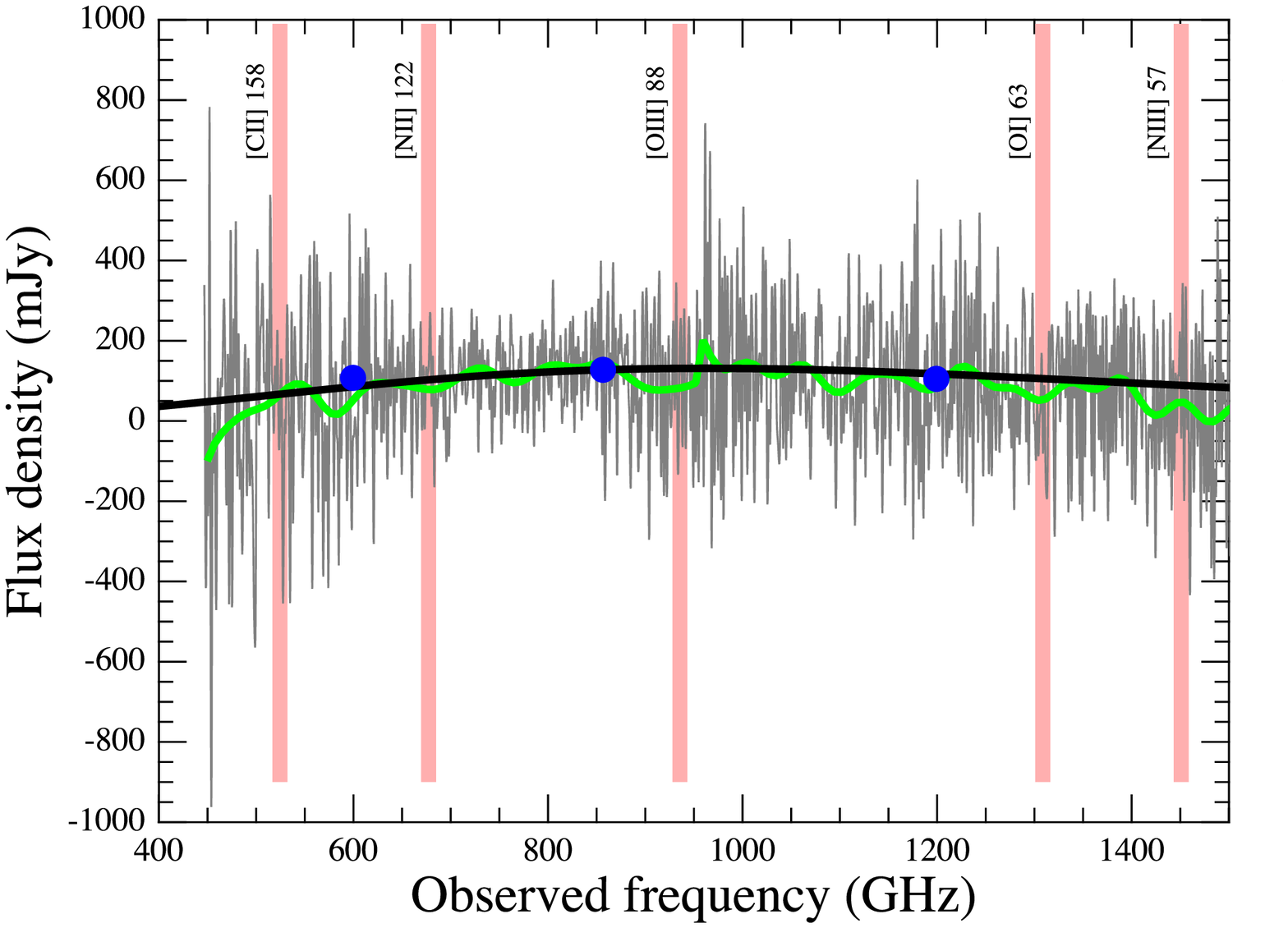}
\caption{The FTS high and low resolution spectra of \hatlas{81} (left) and \hatlas{130} (right) overlaid with the three SPIRE photometric points (dots) and the SED template fit. The photometric errors, even including the 7\% flux calibration uncertainty of the SPIRE Photometer, are smaller than the dots. The expected positions of a number of fine-structure lines are indicated with vertical bars.}
\label{fig_fts_spec}
\end{figure*}

The FTS detector timelines, sampled at 80 Hz, are first cleaned of glitches due to high energy particles hitting the substrate, the bolometers or the readout electronics. The glitch removal utilises a wavelet based method. The fraction of samples identified as glitches was $\sim0.5$\% for the SSW and $\sim0.8$\% for the SLW for the three observations. Next, instrumental effects are removed: electrical cross-talk and time domain phase correction. {Then we convert the bolometer recorded voltage into its temperature using a calibration between the bolometer resistance and temperature, established during ground testing. This is in contrast to the standard voltage-based pipeline FTS processing, where a non-linearity correction is applied to the read-out voltages. The reason for using bolometer net temperature is that it represents much more closely the physical response of the bolometer to incident radiation and therefore ``naturally'' accounts for any non-linearity and drift in that response.  In the circumstances we are faced with here, where we are critically dependent on being able to accurately subtract two large signals to give a very small one, we have found that this, combined with the interactive phase correction steps, gives a small but significant improvement over the current pipeline processing.  It may be that using the temperature based pipeline ends up being unnecessary with further improvements in the voltage based pipeline, but for the time being we prefer this method, to remove any dependency on the drift correction and non-linearity steps which could inject systematic uncertainties into the data.}

{The resulting detector timelines are then converted to interferograms with the use of the spectrometer mirror mechanism (SMEC) timeline that provides the optical path difference (OPD) at each time sample. Thus, at the end of this step we have interferograms of signal versus OPD. The 100 forward and 100 reverse interferograms are then divided into 2$^{18}$ parts, which we shift iteratively in zero-path difference (ZPD) until the interferograms are precisely aligned, removing any residual phase shift introduced by the detectors' temporal response and beamsplitters \citep{spencer10, naylor10}. The interferograms from all 200 scans are then merged, the forward and the reverse ones separately, they are baseline corrected and during the averaging process the data points which are marked as glitches are discarded.  The next steps include gain correction and phase correction of the interferograms, which are further zero padded and converted to the spectral domain using Fast Fourier Transforms.  To convert the flux to physical units (Jy) and to correct for the relative spectral response function (RSRF) we use a deep observation of Uranus. At this stage, the forward and reverse interferograms are averaged together. This procedure was followed for the two targets and the reference sky. The last step is the subtraction of the reference sky from the target and the resulting spectra of \hatlas{81} and \hatlas{130} are shown in Figure~\ref{fig_fts_spec}.}

As we described in the previous section, the absolute continuum calibration, especially for faint targets, is quite difficult. Small changes, of the order of few tens of mK, in the temperature of the instrument and the telescope during the reference dark sky observation and the target observations can lead to quite significant offsets in the continuum level. We iteratively adjusted the overall dark sky level in order to get the best match with the available photometric points (Swinyard et al., in preparation). A further correction was applied to the SLW spectra for both targets in order to remove the effect of variations in instrument temperature \citep{spencer10}. These corrections are already applied to the spectra shown in Figure~\ref{fig_fts_spec}. Note that the continuum offset affects the low resolution continuum too. This makes the use of the FTS in low resolution mode, for an SED-like measurements, not very suitable, especially for faint targets.

%%%%%%%%%%%%%%%%%%%%%%%%%%%%%%
%%%%%%%%%%
\section{Results}
\label{sec_res}

The FTS covers the wavelength region from 197 to 670\,\microns. Some of the most prominent and well studied ISM cooling lines fall in this range, for the redshifts of the two targets in this study. {The FTS spectral resolution is $\Delta\nu = 1.2$ GHz for the high resolution mode and it is constant in frequency space. This spectral resolution corresponds to instrumental line FWHM from 280 to 970 km s$^{-1}$, over the FTS spectral range. The measured FWHM of the CO(1-0) line is $440 \pm 50$ \kms\ \citep{frayer10}, also compatible with the EVLA measurement (see Figure~\ref{fig_co10_radio}). This means that some lines could be marginally resolved at $\lambda < 300$ \microns.

The instrumental line shape of all FTS instruments is a $Sinc$ function due to the truncation of the interferogram by the limited travel of the moving mirror. This function is characterised by one central peak and a number of secondary maxima at lower intensity. There are techniques to suppress the secondary maxima at the price of lowering the resolution, the so called apodization \citep{naylor07}, however we did not perform this additional smoothing as we are looking for very faint features. The SPIRE FTS instrumental line shape is very close to a $Sinc$ function, with noticeable asymmetry of the first high-frequency minimum. The reason for this asymmetry is still under investigation by the SPIRE instrument team. This may have very small effect on the line flux measurements, however the line centroids are not affected, as it was shown that the centroids of CO ladder lines in nearby sources are as good as 1/20 of the resolution element \citep{OM}.

Thanks to the already available high accuracy spectroscopic redshifts for the two galaxies \citep{negrello10,lupu10,frayer10}, we know where to search for lines at specific wavelengths. We fit a $Sinc$ function model to the continuum subtracted spectrum in \mbox{$\pm 5000$ \kms}\ rest-frame, around the expected positions of a number of ISM lines:
\begin{equation}
I(v) = I(0)\,\frac{\sin((v-v_0)/v_\sigma)}{(v-v_0)/v_\sigma}, 
\end{equation}
where $v_{\sigma}$ is the $Sinc$ line width in \kms, measured between the two zero crossings nearest to the peak and $v_0$ is the line centroid offset from the expected position for the given redshift. For the high resolution mode we have \mbox{$v_\sigma = 1.193\,\left(\lambda_c/\mu\mathrm{m}\right)$ \kms} at the line centroid $\lambda_c$, we keep this parameter fixed. Note that the line FWHM is $v_\mathrm{FWHM} = v_\sigma/1.20671$. The output parameters are the line peak $I(0)$ and the line centre offset $v_0$. The integrated line flux for for a $Sinc$ function is then calculated using:
\begin{equation}
S_\mathrm{line} = 0.012\,\left(\frac{I(0)}{\mathrm{mJy}}\right)\times 10^{-18}\, \mathrm{W\, m}^{-2},
\label{eq_flux}
\end{equation}
and the line luminosities with: 
\begin{equation}
L_\mathrm{line} = 3.11\times 10^{7}\, \left(\frac{S_\mathrm{line}}{10^{-18}\,\mathrm{W\, m}^{-2}}\right)\,\left(\frac{D_L}{\mathrm{Gpc}}\right)^2\ L_{\sun},
\label{eq_lumin}
\end{equation}
where $D_L$ is the luminosity distance. 

\subsection{\hatlas{81}}

The most prominent features in the FTS spectrum of \hatlas{81} are the \oiii\  88 \microns\ and \cii\  158 \microns\ lines, clearly seen in Figure~\ref{fig_fts_spec}, left panel. The continuum subtracted spectral segments of \hatlas{81}, at $\pm 5000$ \kms\ rest-frame around both lines, are shown in Figure~\ref{fig_lines}. The derived emission line parameters from the $Sinc$ model fit are given in Table~\ref{tab_lines}.  

{We detect the $^2\mathrm{P}_{3/2}-^2\mathrm{P}_{1/2}$ \cii\  line at 158 \microns, at a signal-to-noise ratio (SNR) of $\sim5$ (see Figure~\ref{fig_lines}, left panel). The line falls near the longer wavelength end of the FTS spectrum, in a region in which the instrument contribution dominates even above the telescope background emission and consequently it is visibly noisier (see Figure~\ref{fig_fts_spec}, left panel). The other prominent feature is the $^3\mathrm{P}_1-^3\mathrm{P}_0$  \oiii\  88\microns\ line, which is the third detection of this line at $z>0.05$ \citep{ferkinhoff10}. We do not detect at a sufficient significance any of the other possible ISM lines that fall in the FTS range. For those lines we give the upper limits in Table~\ref{tab_ulimits}.}

\begin{figure*}
\includegraphics[width=8cm]{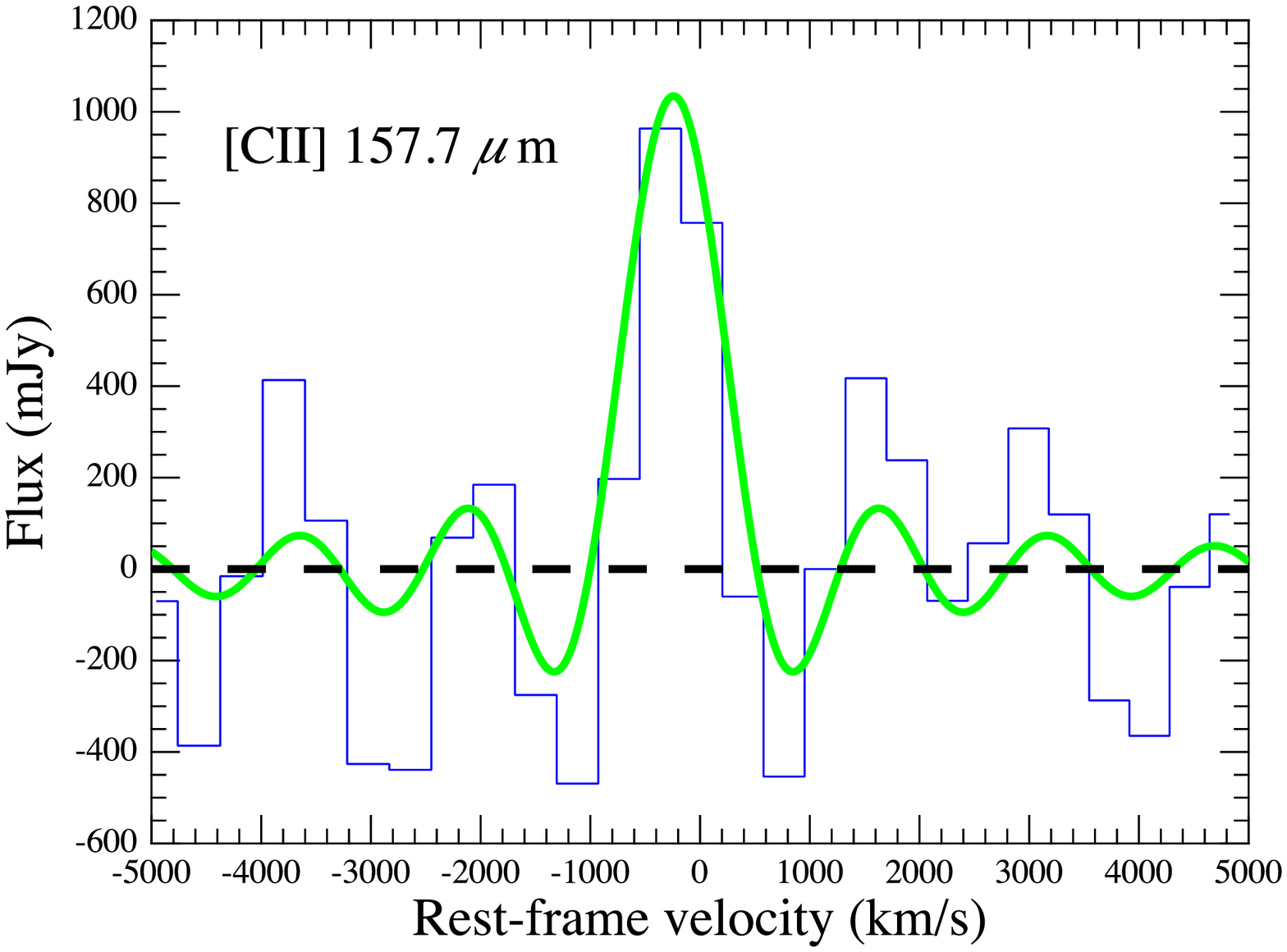}
\includegraphics[width=8cm]{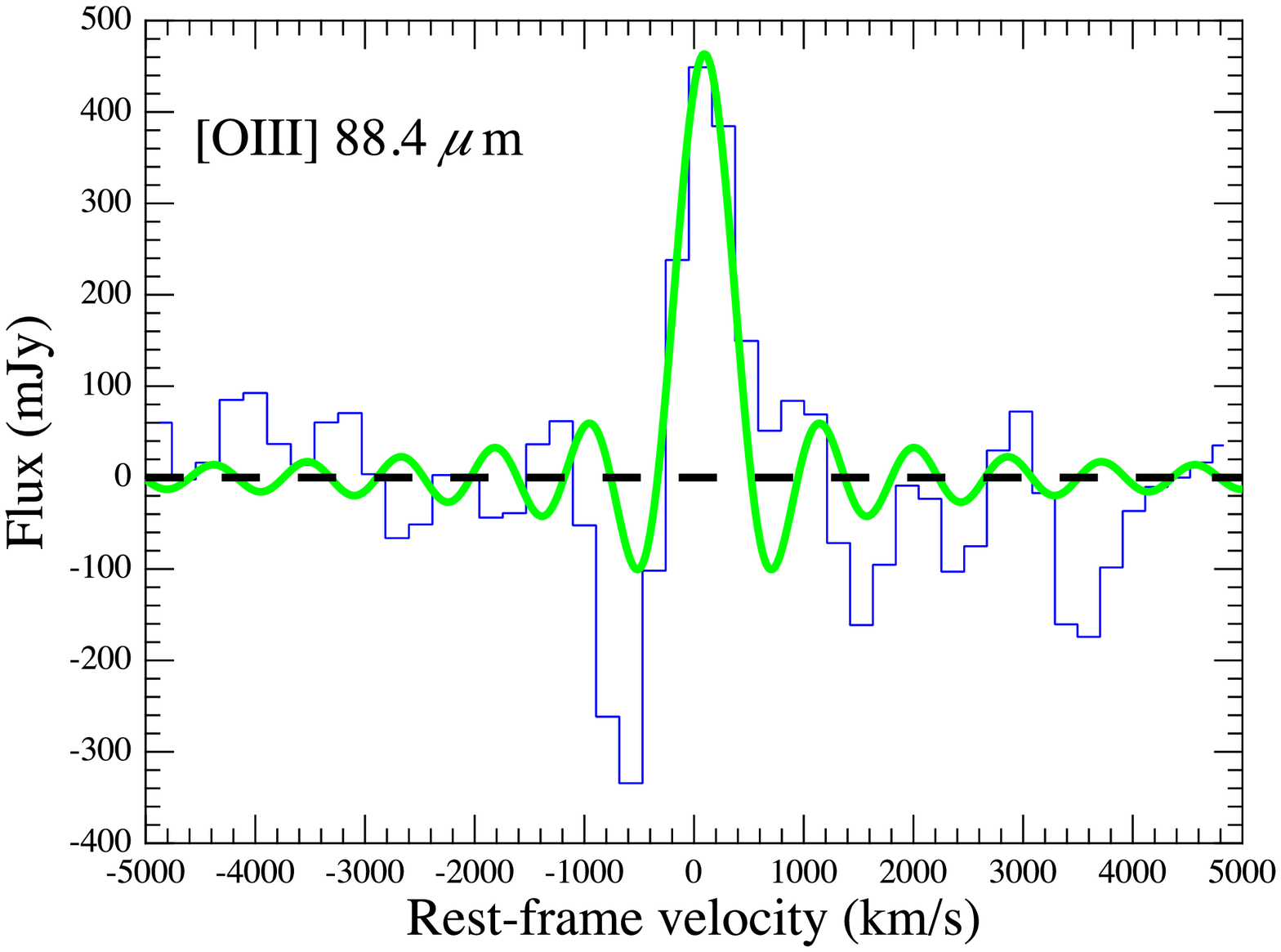}
\caption{Continuum subtracted regions within $\pm 5000$ km s$^{-1}$ rest-frame for \cii\  157.7 \microns\ (left) and \oiii\  88.4\microns\ (right) lines of \hatlas{81}. The FTS data processed using two reference dark sky observations (see Section~\ref{sec_dp}) is shown as histogram. The dashed line indicates the zero level. The best fit $Sinc$ function is also shown.}
\label{fig_lines}
\end{figure*}

\begin{table*}
\caption{Characteristics of the \cii\  and \oiii\  lines in \hatlas{81}, as derived with a $Sinc$ function fit. We use $\mu_\mathrm{L}\,L_\mathrm{IR}(3-2000\mu\mathrm{m}) = (5.13 \pm 0.88)\times10^{13}\,L_{\sun}$ and $\mu_\mathrm{L} = 25 \pm 7$, based on an improved lens model for \hatlas{81}, which is different but still compatible with $\mu_\mathrm{L} = 19$ quoted in \citet{negrello10}. The reported velocity offset $v_0$, \textit{in rest-frame} \kms, is between the fitted line centroid and the expected one for the CO measured redshift $z=3.042$. The instrumental line FWHM in \kms\ corresponds to the high spectral resolution of 1.2 GHz, we keep this parameter fixed for the $Sinc$ fit. The noise \textit{rms} is the standard deviation of the residual (fitted model subtracted from the data) in the velocity range $\pm 5000$ \kms. The reported signal-to-noise ratio (SNR) is the fitted line peak divided by the noise \textit{rms}. The integrated line flux and line luminosity are calculated using Eqs.~\ref{eq_flux} and \ref{eq_lumin} respectively. The errors are statistical errors, they do not include the FTS flux calibration uncertainty, conservatively estimated to be 15-20\% for wavelengths less than 300 \microns\ and 20-30\% for wavelengths above 300 \microns.}
\begin{center}
\begin{tabular}{lll}
\hline\hline
 Characteristics & \oiii\ 88.4 \microns & \cii\ 157.7 \microns \\ \hline
Fitted line centroid $\lambda_c$ (\microns) & $357.2 \pm 0.6$ (839 GHz) & $638.4 \pm 1.8$ (490 GHz) \\
Redshift & $3.041 \pm 0.006$ & $3.046 \pm 0.011$ \\
Velocity offset $v_0$ (\kms) & $95 \pm 30$ & $-240 \pm 60$ \\
Instrumental line FWHM $v_\mathrm{FWHM}$ (\kms) & 514 & 919 \\
Fitted line peak $I(0)$ (mJy) & $480 \pm 70$ & $1100 \pm 160$ \\
Noise \textit{rms} (mJy) & 90 & 211 \\
SNR & 5.4 & 5.2 \\
Line flux $S_{\mathrm{line}}$ ($\times 10^{-18}$ W m$^{-2}$) & $5.8 \pm 0.8$ & $13.2 \pm 1.9$ \\
Velocity integrated line flux $S_{\mathrm{line}}\,\Delta v$ (Jy \kms) & $210 \pm 30$ & $840 \pm 120$ \\
Line luminosity $\mu_\mathrm{L} L_{\mathrm{line}}$ ($\times 10^{10}\ L_{\sun}$) & $13 \pm 2$ & $29 \pm 4$ \\
$L_{\mathrm{line}}/L_\mathrm{IR}$ ($\times 10^{-3}$) & $2.4 \pm 0.5$ & $5.6 \pm 1.3$ \\ \hline
\end{tabular}
\label{tab_lines}
\end{center}
\end{table*}

\subsection{\hatlas{130}}

{We do not have any positive line detection in the FTS spectrum of \hatlas{130}. Even stacking the spectra around all the lines in Tables~\ref{tab_lines} and \ref{tab_ulimits} do not result in any positive signal. This target is about 30\% fainter than \hatlas{81} at 350 \microns\ (see Table~\ref{tab1} and Figure~\ref{fig_fts_spec}). Scaling from \hatlas{81} and assuming the same luminosity of the \cii\  158 \microns\ line, we estimate the expected line flux for \hatlas{130} to be $\sim 4.5\times 10^{-18}$ W m$^{-2}$, taking into account the magnification factor. This is $\sim 3$ times fainter than the measured integrated line flux of \hatlas{81}. Consequently, to detect this line at the same significance would require 9 times longer integration time. On the other hand, the CO(1-0) line luminosity of \hatlas{130} is ~50\% higher than that of \hatlas{81} (see \citealt{frayer10}), so scaling directly with the molecular gas mass, we would expect to see a marginal, $\sim ~3$-$3.5\sigma$, \cii\  158 \microns\ feature. There could be different reasons for a suppressed \cii\  (e.g. \citealt{luhman03}) but also the scaling of the \cii\  with the molecular mass may not be straightforward and tentatively may indicate different conditions of the ISM in \hatlas{130}. }

%{Apart from physical reasons of non-detection, there could be a prosaic one -- a simple pointing offset. The measured \herschel\ absolute pointing error is $2\arcsec$ (68\% confidence interval, \citealt{pilbratt10}). Assuming a Gaussian distribution of the pointings, an offset of $6\arcsec$ (i.e. $3\sigma$) could lead to an overall loss of continuum signal of $\sim 25$\%. It is quite difficult to assess if indeed there was a pointing offset during the observation. If the peak-to-continuum ratio for the lines is constant then this could certainly lead to non-detection, especially for faint targets at the border case where we are looking for features at SNR of $\sim 3$-4.}

Because of the non-detection we have used the observation of \hatlas{130} as a second dark sky measurement, which helped to minimise further the noise in the \hatlas{81} spectrum (see Section~\ref{sec_fts_obs}).

\subsection{Physical characteristics of the emitting region in \hatlas{81}}

We follow the analysis presented in \citet{hailey10} and \citet{stacey10} using the photo-dissociation regions (PDR) models of \citet{kaufman99} with the help of the PDR toolkit.\footnote{The PDR modelling toolkit \citep{pound07} is available at \url{http://dustem.astro.umd.edu}.} The expected physical conditions from the observed lines and the total infrared luminosity are given in terms of PDR cloud density $n$ in cm$^{-3}$ and the FUV ($6 < h\nu < 13.6$ eV) ionization field strength $G_0$, in units of the Habing field -- the local Galactic interstellar radiation field of $1.6\times10^{-3}$ erg cm$^{-2}$ s$^{-1}$. We must emphasise that the results we derive are for a particular set of PDR models with pre-defined parameters, like the visual extinction of $A_V=10$ and metallicity equal to the local ISM metallicity, i.e. $Z = Z_{\sun}$. At much lower metallicities, i.e. $Z = 0.1Z_{\sun}$ and smaller $A_V$ we would expect that the derived $n$ and $G_0$ will change by $\sim10$\% \citep{kaufman99}. In addition, as \hatlas{81} is unresolved, the derived PDR characteristics are an ensemble average of possibly many photo-dissociation regions falling in the SPIRE-FTS beam. This has an implication for the comparison with the PDR models of \citet{kaufman99} using optically thick lines, like the CO transitions or the \oi\ 63\microns\ line. {The CO(1-0) emission in SMGs typically comes from a relatively smooth, optically thick medium (e.g. \citealt{carilli10}) with some substructure that may become more prominent in higher-J lines and the continuum. For such a medium, CO observations could probe higher by a factor of 2 column densities. Nonetheless, in the subsequent analysis, following \citet{kaufman99} and other works (e.g. \citealt{swinbank10,danielson11}), we double the observed luminosities of the optically thick lines, in order to account for the fact that we see the emission only from one side of the clouds.} 

\subsubsection{\cii\  158 $\mu$m line}

{The \cii\  line at 158 \microns\ is one of the best studied ISM collisionally excited lines. It is a widespread coolant for the ISM in galaxies; it requires only relatively modest FUV fluxes to be produced, and hence it traces star formation rather than AGN activity. This line can account for up to 1\% of the total far-IR luminosity in normal galaxies (e.g. \citealt{stacey91}) and is optically thin in most cases. A large fraction of the \cii\  luminosity comes from photo-dissociation regions, however a non-negligible fraction of the \cii\  emission may arise from diffuse ionised gas and from \hii\ regions. The usual approach to estimate this fraction is with the \nii\,122\,\microns\ line (see e.g. \citealt{malhotra01,contursi02,oberst06,vasta10}). Nitrogen has ionization potential 14.53 eV, higher than the 13.6 eV of Hydrogen, and so the \nii\,122 \microns\ line can only originate from \hii\ regions or from the diffuse ionized medium.  When there is not enough information from other lines, which is the case for most of the high redshift submm galaxies, it is assumed that 70\% of the \cii\  emission arises in PDR (e.g. \citealt{hailey10, stacey10}) based on \citet{oberst06} measurements in our Galaxy (the Great Carina nebula) and also supported by observations of nearby normal galaxies \citep{malhotra01}. We follow the same approach and adopt the same 70\% fraction of \cii\  coming from a PDR, which is consistent with the observed lower limit of the \cii/\nii\ ratio of 2.6 in \hatlas{81}, which, compared to the ratio of 1.6 from a pure H\,II region, means that at least 40\% of the \cii\  flux comes from PDRs.}

The ionising UV source can be either hot stars or an AGN and this increases the complexity when one wants to derive the physical conditions based on cooling lines in the FIR. A line ratio, involving \cii, that can be used to infer the origin of the radiation field, is $L_\mathrm{[OI] 63\mu}/L_\mathrm{[CII]}$ \citep{abel09,meijerink07}. The upper limit of this ratio for \hatlas{81} is 1.1, which is characteristic for environments excited by stellar radiation, but also overlaps marginally with the allowed region for AGN excitation, which would require a line ratio greater than 1 \citep{abel09}. Thus, we have to keep in mind that part of the ionizing radiation field may have contribution from an AGN. The FIR lines are not sensitive indicator of an AGN (see e.g. \citealt{gracia11}) which makes it difficult to quantify this fraction. In the subsequent PDR analysis we assume that the radiation field is due to UV coming from hot stars.

%\begin{figure*}
%\includegraphics[width=16cm]{images/Stacey-ciifir.pdf}
%\caption{Diagnostic diagram of $L_\mathrm{[CII]}/L_\mathrm{FIR}$ as a function of $L_\mathrm{FIR}$ for \hatlas{81} (red filled circle) and \hatlas{130} (orange filled circle with an arrow), compared to local and other known high redshift galaxies of different classes \citep{stacey10}. The \cii\  luminosity was corrected for the fraction of the flux arising from H\,II and diffuse ionized regions.}
%\label{fig_cii_fir}
%\end{figure*}

\begin{table}
\caption{Upper limits ($3\sigma$) for some ISM lines for \hatlas{81} that fall in the FTS spectral range. The upper limits were calculated using Eq.~\ref{eq_flux}, assuming $I(0)$ to be $3\sigma$ of the noise st.dev. within $\pm5000$ \kms\ around the expected line centre.}
\begin{tabular}{lrccc}
\hline\hline
Line ID & $\lambda_0$ & $S_{\mathrm{line}}$ & $\mu_\mathrm{L} L_{\mathrm{line}}$ & $L_{\mathrm{line}}/L_\mathrm{IR}$ \\
       & (\microns) & ($10^{-18}$ W m$^{-2}$) & ($10^{10}L_{\sun}$) & ($\times 10^{-3}$) \\ \hline
\oiii & 52.0 &  $<4$ & $<9$ & $<1.7$\\
\textrm{[N\,{\sc iii}]} & 57.0 &  $<4$ & $<10$ & $<1.9$ \\
\oi  & 63.2  &  $<5$ & $<11$ & $<2.2$\\
\nii & 122.1 &  $<5$ & $<11$ & $<2.2$\\
\oi & 145.5  &  $<8$ & $<19$ & $<3.6$\\
\hline
\end{tabular}
\label{tab_ulimits}
\end{table}

An idea of the physical conditions, in terms of the FUV radiation field strength $G_0$ and PDR density $n$, can be obtained by comparing the fractional \cii\  luminosity versus the fractional CO(1-0) luminosity with PDR models \citep{stacey91,hailey10,ivison10d,stacey10}. In the diagram shown in Figure~\ref{fig_c2_co}, where we have used the EVLA CO(1-0) measurements from this work, we can see that even with the \cii\ alone \hatlas{81} is comprised of PDRs with a mean density of $\sim 10^2-10^4$ cm$^{-3}$ and $G_0 \sim 10^2-10^3$, similar to other star-forming galaxies at $z=1-2$ and to SMMJ2135 \citep{ivison10d}. We can compare the relative strength of the \cii\  line, in terms of the FIR luminosity, with other galaxies with measured \cii\ lines (see e.g. \citealt{maiolino09,stacey10,gracia11}). \hatlas{81} has $\log\left(L_{\mathrm{[CII]}}/L_{\mathrm{FIR}}\right) = -2.2 \pm 0.05$ -- one of the highest observed ratios, which contrasts with the deficiency reported for ULIRG and AGN (e.g. \citealt{luhman03}). Although it might seem that \hatlas{81} relative \cii\ strength is much higher than the one seen in AGNs at $z=1-2$, recent measurements with the \herschel\ PACS instrument, however, do not show any clear separation between local AGN systems (QSO, Seyferts, LINERS), starbursts and infrared luminous galaxies \citep{gracia11}. 

\begin{figure*}
\includegraphics[width=15cm]{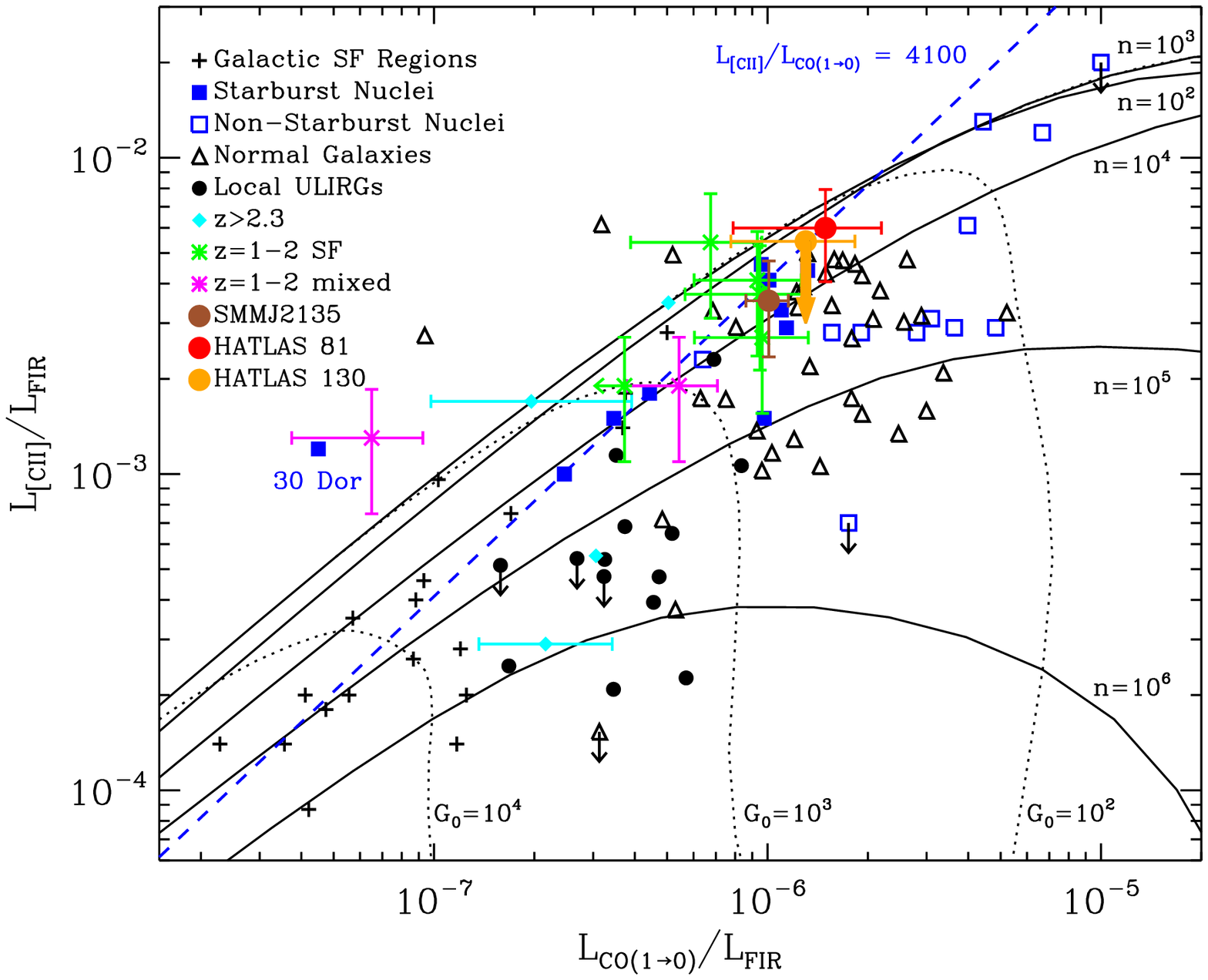}
\caption{Diagnostic diagram of $L_\mathrm{[CII]}/L_\mathrm{FIR}$ as a function of $L_\mathrm{CO(1-0)}/L_\mathrm{FIR}$ for \hatlas{81} (the rightmost filled circle) and \hatlas{130} (filled circle with a down arrow), compared to local and other known high redshift galaxies of different classes \citep{stacey91,stacey10,hailey10}. The lines with constant density $n$ and constant $G_0$ from PDR models of \citet{kaufman99} are also shown. The \cii\  luminosity was corrected for the fraction of the flux arising from \hii\ and diffuse ionized regions. The CO(1-0) line flux from the EVLA measurement was multiplied by 2 to take into account that the line is optically thick and we see only the fraction coming from the illuminated PDR side. \lines{[CII]}/\lines{CO(1-0)} of 4100, typical for local star-burst galaxies, is also shown as a dashed line. We have used $L_\mathrm{FIR}(42.5-122.5\mu\mathrm{m}) = L_\mathrm{IR}(3-2000\mu\mathrm{m})/1.54$, derived by integrating the SED of \hatlas{81}. }
\label{fig_c2_co}
\end{figure*}

The ratio \lines{[CII]}/\lines{CO(1-0)} of $\sim 4000$ is quite close to the average of 4100 for normal metallicity star-forming galaxies or Galactic star-formation or molecular clouds \citep{madden00,stacey10}. It is also important to note that in comparison with the other high redshift galaxies \hatlas{81} has one of the highest fractional \cii\ and CO(1-0) luminosities. Whether this is significant or it is simply due to luminosity bias is to be confirmed by increasing the number of high redshift SMGs at $L_\mathrm{FIR} < 10^{12}\, L_{\sun}$. 

The diagram shown in Figures \ref{fig_c2_co} is quite useful as diagnostic and as a comparison with other observations of galaxies of different classes. To properly quantify the physical conditions, we can add all the available information: the PDR luminosity of \cii, the total FIR luminosity, the upper limit of the \oi\ 63\microns\ line and the EVLA detections of CO(1-0) and perform a least square fit with grids of PDR models with different $(n,G_0)$. Using the PDR toolkit \citep{pound07} we derive the best fit PDR density of $n = 10^{3.30\pm 0.22}$ cm$^{-3}$, which is in very good agreement with the H$_2$ density from RADEX models using the CO lines (\citealt{lupu10}, model m03). The corresponding radiation field strength is $G_0 =10^{2.24 \pm 0.19}$. The confidence $\chi^2$ region in the $(n,G_0)$ plane, delineated by the line ratios, is shown in Figure~\ref{fig_n_g0}. These $(n,G_0)$ values correspond to a PDR surface temperature of $T_s \approx 200$ K. 

Adding the CO(7-6) line from \citet{lupu10}, and considering only the two ratios \cii/CO(1-0) and CO(7-6)/CO(1-0), we obtain an order of magnitude higher values: $n \approx 3.2\times 10^{4}$ cm$^{-3}$ and $G_0 \approx 5.6\times 10^3$, as shown in Figure~\ref{fig_n_g0}. This is to be expected as the CO(7-6) line usually traces higher density regions. In addition, it must be noted that CO(7-6) is likely to be blended with [C\,{\sc i}] given the relatively coarse resolution of the Z-Spec spectrometer. Recent measurements  with \herschel\ show that [C\,{\sc i}] has approximately 50-150\% of the flux of the CO(7-6) line, 80\% on average, based on a sample of 20 local ultra-luminous infrared galaxies (van der Werf et al, in preparation). Even taking into account the blending, i.e. scaling down the CO(7-6) line flux by a factor of 1.8, the overall picture does not change significantly. It should be noted that $G_0$ derived from \cii\ should be more representative for the PDR as a whole, since the \cii\ line comes from PDRs \textit{and} molecular cloud surfaces. Consequently, in the following analysis we use the low $(n,G_0)$ values (i.e. not taking into account the CO(7-6) line).

A straightforward use of the PDR derived characteristics is to investigate the effective physical size of the non-lensed emitting region. With PDR density $n$ of $\approx 2000$ cm$^{-3}$ and the total gas mass from \citet{lupu10} of $M_{gas} = 1.3\times10^{11} M_{\sun}/\mu_\mathrm{L}$, where $\mu_\mathrm{L}$ is the magnification factor of $25\pm7$, we obtain the effective radius of the PDR emitting region of $\sim 1.0\ \mathrm{kpc}/\mu_\mathrm{L}^{1/3}$ or (0.3-0.4) kpc for a spherical distribution. For a disk geometry with thickness of 100 pc, the radius is $\sim 3.0\ \mathrm{kpc}/\sqrt{\mu_\mathrm{L}}$ or (0.5-0.7) kpc correspondingly. This result is in good agreement with the equivalent dust emitting radius of 0.7 kpc for \hatlas{81} derived in \citet{lupu10}, and the size of the emitting region in SMMJ2135 \citep{danielson11}.

Although the SPIRE FTS smallest beam size is $17\arcsec$ FWHM, nevertheless we can have an idea of the the effective magnified size of the \cii\ emitting region, by means of the \cii\ beam filling factor. Assuming that the PDR slab is illuminated from one side by a flux $G_0$ and that dust grains are heated by the FUV photons then we can compare the model predicted \cii\  line intensity \citep{kaufman99} with the observed line intensity and thus we can derive the beam filling factor \citep{wolfire90} $\Phi_\mathrm{A} = \left(S_{\mathrm{[CII]}}/\Omega\right)/I_{\mathrm{[CII]}} \approx 3\times10^{-4}$, where $\Omega = 10^{-7}\ \mathrm{sr} = 1.22$ arcmin$^2$ is the SPIRE FTS effective beam area at the wavelength of the \cii\  line \citep{swinyard10}. This corresponds to an effective radius of the magnified emitting region of $\approx 1\arcsec$. This is consistent with the SMA continuum observations and the EVLA CO(1-0) map, shown in Figure~\ref{fig_co10_radio}.
%Following \citet{wolfire90}, and assuming that a central source is responsible for the $G_0$ illumination over a uniform cloud distribution out to a distance $R$, then substituting $G_0$ of 300 from the PDR analysis we derive the total size of the galaxy to be of the order of 14 kpc.

We can also use the PDR results to place a lower limit on the H$_2$ mass of \hatlas{81}, independently from the mass derived using CO measurements. Assuming that the \cii\  line is optically thin then the minimum total atomic hydrogen mass in the PDR \citep{crawford85,hailey10,ivison10d}:
\begin{equation}
M_\mathrm{a} = 0.77 \, Q(T)\, \frac{L_\mathrm{[CII]}}{L_{\sun}}\, \frac{1.4\times10^{-4}}{\chi(\mathrm{C}^{+})}\, M_{\sun},
\end{equation}
where $Q(T)=\left[1 + 2e^{(-91\,K/T_s)} + n_{crit}/n\right]/\left[2e^{(-91\,K/T_s)}\right]$. Substituting the \cii\  luminosity corrected for the \hii\  contribution and for the magnification, the PDR surface temperature and density, the critical density of C$^{+}$ $^{2}P_{2/3}$ level of $n_{crit} = 2700$ cm$^{-3}$ \citep{launay77} and the C$^{+}$ abundance $\chi(\mathrm{C}^{+})=1.4\times10^{-4}$ \citep{savage96}, we derive $M_\mathrm{a} \approx 1.7\times10^{10} M_{\sun}$.  It should be noted that this minimum atomic mass in the PDR is quite sensitive to the assumed PDR fraction of the \cii\  luminosity, the magnification factor and the unknown real heavy element abundances. Nevertheless the derived $M_\mathrm{a}$ is of the order of the estimated molecular gas mass $M(\mathrm{H}_2) = (1-3)\times 10^{10}\, M_{\sun}$ derived from CO(1-0) \citep{frayer10}, somewhat higher than the fraction of 20-50\% seen in other high-redshift star-forming galaxies \citep{stacey10}.

\subsubsection{\oiii\  88 $\mu$m line}

The \oiii\  lines are usually connected with \hii\ regions but also with AGN activity. The \oiii\  88\microns\ line is emitted via the transition $^3P_1-^3P_0$ of O$^{++}$. The ionization potential of O$^{+}$ is 35 eV, which requires a nearby hot young star or stars  ($T \sim 36\,000-40\,000$ K) or an AGN, to provide the necessary FUV photons. The \oiii\  line emission is quite sensitive to the effective temperature of the ionizing stars and if this temperature drops from 36,000 K to 33,000 K the line intensity would drop by a factor of up to 30 \citep{rubin85}. The ratio $L_\mathrm{[OIII]}/L_\mathrm{FIR}$ observed in \hatlas{81} is significantly higher than the one observed in normal galaxies, although smaller than the ratio in irregular, \hii\ dominated galaxies \citep{malhotra01,gracia11}. Comparing this ratio with models \citep{abel09} indicate that some fraction of the emission may be due to an AGN source. The situation is not that clear, however, comparing $L_\mathrm{[OIII]}/L_\mathrm{FIR}$ with recent observations, which show no clear separation of the local AGN sources (Seyferts, QSOs, LINERs) and the rest \citep{gracia11}. Thus, similar to the $L_\mathrm{[OI]\,63\mu}/L_\mathrm{[C\,II]}$ result, we cannot discard completely that some fraction of the ionizing field may arise in an AGN.

\begin{figure}
\includegraphics[width=9cm]{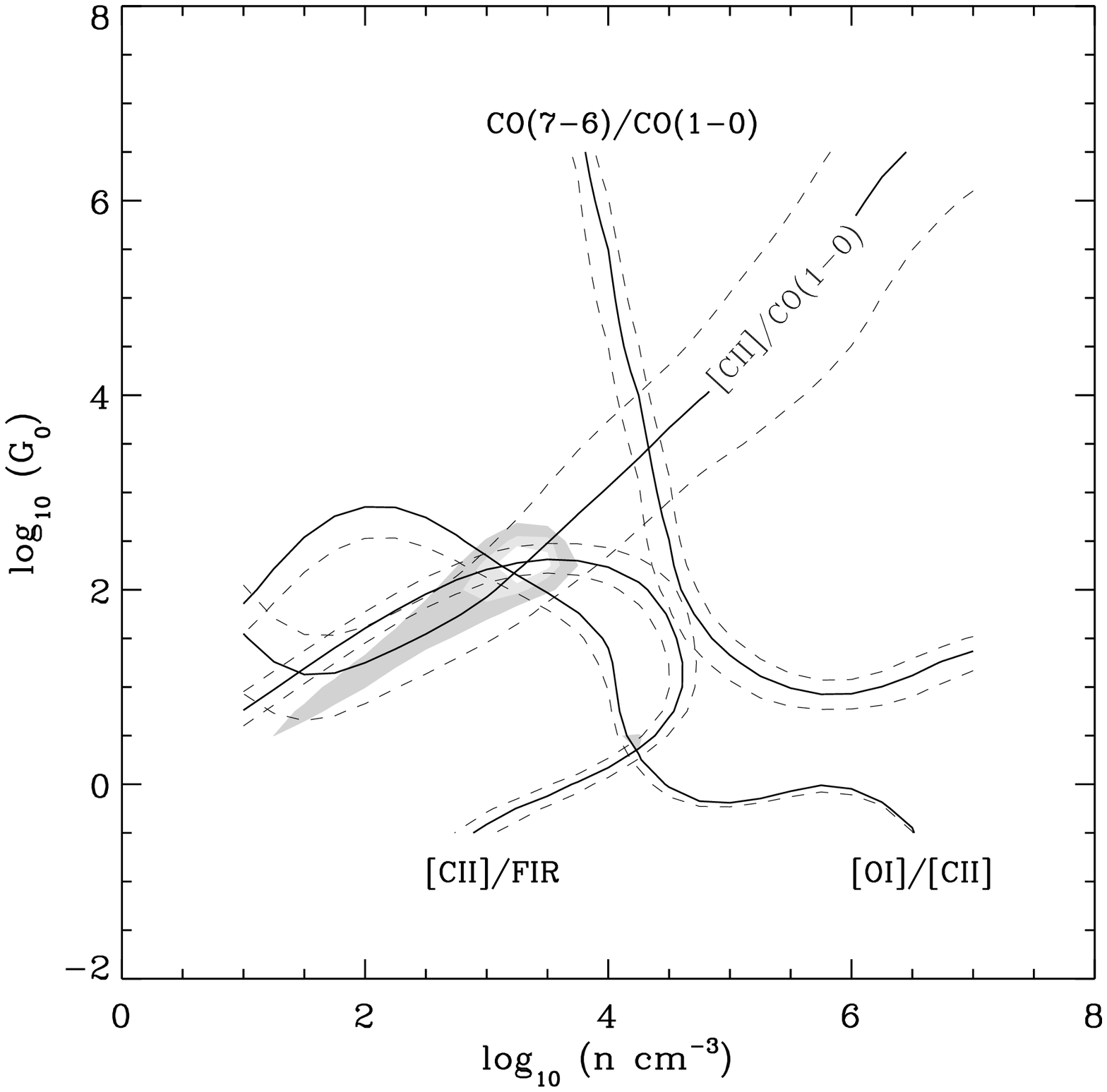}
\caption{PDR model predictions for the density ($n$) and the FUV flux ($G_0$) from the observed line ratios. The \oi\ 63\microns/\cii\ 158 \microns\ is shown as an upper limit, the dashed line indicating the side where we expect the ratio to be if \oi\ is less luminous. The shaded contours show the $3\sigma$ confidence region of the expected $(n, G_0)$ values when considering only \cii, CO(1-0) and the upper limit on \oi. The CO(7-6)/CO(1-0) ratio was corrected for [C\,{\sc i}] blending with CO(7-6).}
\label{fig_n_g0}
\end{figure}

The detected \oiii\  88 \microns\ line can also provide us with information about the properties of {an ensemble \hii\ component, with \oiii\  emission coming from regions much closer to the UV sources} (see e.g. \citealt{cormier10,okada10}). Transitions of the same ion O$^{++}$ from different collisionally excited levels of the same fine-structure multiplet, as is the case for the two lines of \oiii\  at 52 and 88 \microns, are not sensitive to the temperature, for a reasonably large range of the density, and can be used to estimate the electron density $n_e$ in the \hii\ region. Unfortunately we do not detect \oiii\  at 52 \microns, thus we can only put an upper limit on $L_{\mathrm{[OIII]\,52\mu}}/L_{\mathrm{[OIII]\,88\mu}}$ of $\lesssim 0.7$, which is quite close to the low density limit of $n_e \approx 100 $ cm$^{-3}$ (see e.g. \citealt{rubin89,fischer99}). This electron density is typical for galactic \hii\ regions.

%Using the \oiii line we can also estimate the minimum mass of the ionized gas, assuming that all oxygen in the H\,II region is of O$^{++}$ form and that the temperature and the density are high then we have (see e.g. \citealt{ferkinhoff10}):
%\begin{equation}
%M(\mathrm{H}^{+})  = 0.03\,\frac{L_{\mathrm{[OIII]}}}{L_{\sun}}\, \frac{5.9\times10^{-4}}{\chi(\mathrm{O}^{++})}\, M_{\sun},
%\end{equation}
%where $\chi(\mathrm{O}^{++})$ is the relative O$^{++}$ abundance. Assuming $\chi(\mathrm{O}^{++}) = 5.9\times10^{-4}$ \citep{savage96} and correcting the line luminosity for the lensing amplification we derive the minimum mass of the ionized gas $M(\mathrm{H}^{+}) \geq 1.5\times10^{8}\, M_{\sun}$. This means that the ionized gas is $\sim1$\% of the total molecular gas $M(\mathrm{H}_2)$ in \hatlas{81}.

\subsection{FIR/radio correlation and millimeter imaging source constraints}

In support for the hypothesis that some fraction of the ionizing radiation in \hatlas{81} may originate from an AGN is the radio data (see Section~\ref{evla}). In Figure~\ref{fig_sed_radio} we compare the Spectral Energy Distribution of \hatlas{81} to the template used in \citet{ivison10d} for the SMMJ2135 (the cosmic ``Eyelash''), which follows the FIR/radio correlation. Combining $S_\mathrm{8.4GHz}$ with $S_\mathrm{1.4GHz}=950\pm 150$\,$\mu$Jy from the FIRST survey \citep{becker95} yields a spectral index of $\alpha=-0.17^{+0.11}_{-0.09}$ (where $S_{\nu} \propto \nu^{\alpha}$), suggesting that an AGN is powering some fraction of the radio emission \citep{ibar10a}. Note however that a flat spectrum could also be a result from compact star-forming regions dominated by star formation \citep{hunt05}. Combining the radio measurements with $L_{\rm IR}$ yields $q_{\rm IR}=1.8\pm 0.1$ (see \citealt{ivison10c} for details). This is only barely consistent with the FIR--radio correlation, providing more evidence that \hatlas{81} is moderately radio-loud (cf.\ $q_{\rm IR}=2.4$ for FIR-selected galaxies -- see \citealt{ivison10a,ivison10c,ivison10d,jarvis10,michalowski10}).

\begin{figure}
\includegraphics[height=8cm,angle=-90]{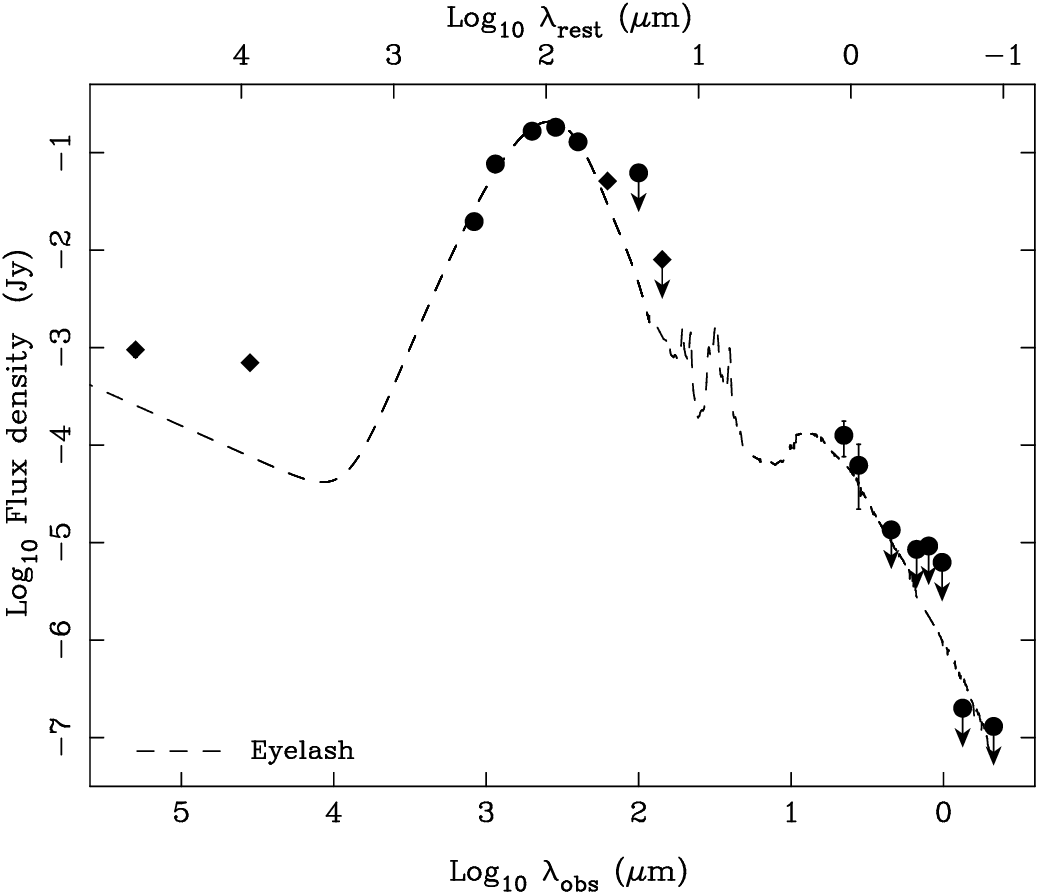}
\caption{SED of \hatlas{81} including the new measurements presented in this paper: the radio observation at 8.5 GHz (see Section~\ref{evla}) and PACS 70 and 160 \microns, all marked as diamonds. The 1.4 GHz detection of \hatlas{81} in the FIRST survey \citep{becker95} is also shown. The dashed line is an Arp~220 template, as used for SMMJ2135 source (the cosmic ``Eyelash'') reported in \citet{ivison10d} and that obeys the FIR/radio correlation. The filled circles represent data from \citet{negrello10} and \citet{hopwood10}. The two radio measurements are clearly deviating from the FIR/radio correlation followed by other FIR selected galaxies.}
\label{fig_sed_radio}
\end{figure}

At mm-band, the peak and the total 1\,mm fluxes measured in a PdBI $0.62\arcsec \times 0.30\arcsec$ beam and $\sim 3\arcsec$ diameter area of \hatlas{81} are 2.4\,mJy and 7.5\,mJy respectively (Neri et al. in preparation), suggesting that the contribution of any strong compact ($<0.54\arcsec$) source, like an AGN -- if it is indeed present in this object, would be at most 33\% of the total mm radiation. If the unresolved 20 mJy flux detected by the IRAM 30 m telescope at 2 mm is also considered \citep{negrello10}, then the PdBI measured peak suggests the contribution of any strong compact source would be less than 14\% of the total flux at these wavelengths. 

With higher resolution observations using CARMA in its most extended configuration, \hatlas{81} is not detected. The \textit{rms} noise in the $0.17\arcsec \times 0.14\arcsec$ beam map is estimated at 0.67\,mJy. Based on this and the peak emission in the PdBI observations, we conclude that if the peak emission in the PdBI image was a point source, it would have been detected; therefore, it has to be at least $0.15\arcsec$ to be consistent with the CARMA non-detection. This implies the 2.4\,mJy peak emission measured in the PdBI beam cannot all come from a point source or AGN, putting an upper limit of the AGN flux contribution to no more than $\sim 1/3$ (or 0.8\,mJy) of the 2.4\,mJy peak and $\sim 1/10$ of the 7.5\,mJy total mm emission detected by PdBI or $\sim 1/25$ of the 20\,mJy detected by the IRAM telescope. This is consistent with the continuum upper limit of 1\,mJy measured at 1\,cm in the EVLA observations, when extrapolated to the 1\,mm wavelength regime for a radio-loud or flat-spectrum source.

Putting all the pieces together we can see that the physical conditions in \hatlas{81} are similar to other high redshift star-forming galaxies, although \hatlas{81} is at somewhat lower intrinsic luminosity, but still being an ultra-luminous infrared galaxy. The radio observations provide supporting evidence to the hints from the ISM lines, that part the ionization field in the galaxy may be due to an AGN. With the available observations we cannot put any strong constraint on this fraction, although a compact source contribution in the mm-band flux cannot exceed 33\%.

\section{Conclusions}
\label{sec_conclusions}

We have presented \herschel-SPIRE Fourier Transform Spectrometer and radio follow-up observations of \herschel-ATLAS strongly lensed sources. Two targets, \hatlas{81} and \hatlas{130} were observed with the FTS, resulting in spectra over a wavelength range from 197 to 670 \microns\ (447-1522 GHz) at a constant spectral resolution of 1.2 GHz. We also presented observations of \hatlas{81} with the EVLA, PdBI and CARMA in the radio domain, helping to better characterise the source properties.

Thanks to the lensing effect, boosting the luminosity of \hatlas{81} by a factor of $\sim 25$, we were able to detect \cii\  at 158 \microns\ and \oiii\  88 \microns\ lines in its FTS spectra. We do not detect any plausible lines in the spectrum of \hatlas{130}. We should have been able to marginally detect the \cii\ 158 \microns\ line at 3-3.5$\sigma$ if we scale the \cii\ with the molecular mass derived from ground-based CO(1-0) line measurements. If we assume the same ratio of \cii\ to $L_\mathrm{IR}$ as in \hatlas{81}, however, we do not expect detection -- the line would be too faint. This is an indication of a possible difference in the quantity  and the distribution of the molecular material in the two FTS targets, although the uncertainties are too important for any strong claim. 

The successful line detections in \hatlas{81} allowed for detailed studies of the physical conditions in the emitting regions at $z \sim 3$. The clear and unambiguous detections of \oiii\  88\microns\ and \cii\  158 \microns\ allow to measure the redshift using both lines. The derived value of $z = 3.043 \pm 0.012$, using the two lines, confirms the measured redshift of $z=3.042$ from ground based CO measurements. The detection of the \oiii\  88 \microns\ line is the third at redshifts higher than 0.05. We expect that the on-going SPIRE and PACS spectroscopy with \herschel\ will lead to many more such line detections.
%It is important to note however that without the precise spectroscopically measured redshift the tentative detections of many of the ISM lines would have been extremely difficult due to the faintness of the object, SED peak flux density of $\sim 200$ mJy, and the noise properties and sensitivity limits of the FTS instrument. 

Combining the FTS detected \cii\,158 \microns\ line, the total infrared luminosity, the upper limit on the \oi\,63\,\microns\ line, the ground based detections of the CO(1-0) transition we were able to infer a PDR cloud-ensemble density $n_0 \approx 2000$ cm$^{-3}$ and the far-UV ionizing field strength $G_0 \approx 200$ in \hatlas{81}. The PDR surface temperature is about 200 K and the effective radius of the PDR emitting region is of the order of 500-700 pc for a disk of 100 pc thickness. These characteristics are similar to other high redshift star-forming galaxies.  

The presence of a strong \oiii\  line, the upper limit of $L_\mathrm{[OI] 63\mu}/L_\mathrm{[CII]}$ and the significant deviation from the FIR/radio correlation are all plausible indications that some fraction of the emission from \hatlas{81} may be due to an AGN source. The conservative upper limit of this contribution in the radio domain is estimated at $\sim33$ \%.

The analysis presented in this paper is a clear illustration of the necessity of a combined effort from follow-up observations at different wavelengths in order to have a better idea of the physical conditions in the star-forming galaxies in the distant Universe. \herschel\ will be indispensable in this task, providing large samples of such galaxies for follow-up studies.

\section*{Acknowledgments}
We are indebted to S. Hailey-Dunsheath for making available his script and the data used to produce some of the figures, as well as for his valuable help with some of the interpretations. We thank Magda Vasta and Serena Viti for useful discussions.\\
The \herschel-ATLAS is a project with \herschel, which is an ESA space observatory with science instruments provided by European-led Principal Investigator consortia and with important participation from NASA. The H-ATLAS website is \url{http://www.h-atlas.org/}. PACS has been developed by a consortium of institutes led by MPE (Germany) and including UVIE (Austria); KU Leuven, CSL, IMEC (Belgium); CEA, LAM (France); MPIA (Germany); INAF-IFSI/OAA/OAP/OAT, LENS, SISSA (Italy); IAC (Spain). This development has been supported by the funding agencies BMVIT (Austria), ESA-PRODEX (Belgium), CEA/CNES (France), DLR (Germany), ASI/INAF (Italy), and CICYT/MCYT (Spain). SPIRE has been developed by a consortium of institutes led by Cardiff University (UK) and including Univ. Lethbridge (Canada); NAOC (China); CEA, LAM (France); IFSI, Univ. Padua (Italy); IAC (Spain); Stockholm Observatory (Sweden); Imperial College London, RAL, UCL-MSSL, UKATC, Univ. Sussex (UK); and Caltech, JPL, NHSC, Univ. Colorado (USA). This development has been supported by national funding agencies: CSA (Canada); NAOC (China); CEA, CNES, CNRS (France); ASI (Italy); MCINN (Spain); SNSB (Sweden); STFC (UK); and NASA (USA).

\label{lastpage}
\end{document}